# Flexible uniform-sampling foveated Fourier single-pixel imaging

HUAN CUI,[1, †] JIE CAO, [1, 2, †,*] QUN HAO,[1,3,*] HAOYU ZHANG,[1] CHANG ZHOU [1]

[1] *Key Laboratory of Biomimetic Robots and Systems, Ministry of Education, Beijing Institute of Technology, Beijing, 100081, China*
[2] *Yangtze Delta Region Academy, Beijing Institute of Technology, Jiaxing 314003, China*
[3] *Changchun University of Science and Technology, Changchun, 130022, China4*
[†]*The authors contributed equally to this work.*
*Corresponding author: ajieanyyn@163.com; qhao@bit.edu.cn*

**Abstract: Fourier single-pixel imaging (FSI) is a data-efficient single-pixel imaging (SPI). However, there is still a serious challenge to obtain higher imaging quality using fewer measurements, which limits the development of real-time SPI. In this work, a uniform-sampling foveated FSI (UFFSI) is proposed with three features, uniform sampling, effective sampling and flexible fovea, to achieve under-sampling high-efficiency and high-quality SPI, even in a large-scale scene. First, by flexibly using the three proposed foveated pattern structures, data redundancy is reduced significantly to only require high resolution (HR) on regions of interest (ROIs), which radically reduces the need of total data number. Next, by the non-uniform weight distribution processing, non-uniform spatial sampling is transformed into uniform sampling, then the fast Fourier transform is used accurately and directly to obtain under-sampling high imaging quality with further reduced measurements. At a sampling ratio of 0.0084 referring to HR FSI with 1024×768 pixels, experimentally, by UFFSI with 255×341 cells of 89% reduction in data redundancy, the ROI has a significantly better imaging quality to meet imaging needs. We hope this work can provide a breakthrough for future real-time SPI.**

## 1. Introduction

Unlike the traditional imaging via a pixelated detector array, single-pixel imaging (SPI) can reconstruct object information by means of light intensity correlation measurements via a single-pixel detector with no spatial resolution [1-3]. Due to the advantages of high sensitivity and wide spectrum range, SPI has been considered for using in many fields, especially where traditional imaging methods are costly or technically unavailable, such as infrared imaging [4], X-ray imaging [5], and terahertz imaging [6].

Fourier single-pixel imaging (FSI) is a data-efficient SPI scheme [7]. FSI uses orthogonal Fourier basis patterns to obtain the target spectral information and then directly use the inverse fast Fourier transform (IFT) to reconstruct the target. With the advantage of sparse spectrum property and fast reconstructing calculation, both the imaging quality and the reconstruction efficiency of under-sampling FSI are better than other SPI schemes. However, there is still a serious challenge on low-sampling FSI to obtain higher imaging quality using fewer measurements, because of the ringing artifact and detail loss [8]. Many methods have been proposed to further enhance the imaging quality of low-sampling FSI, such as optimizing the sampling strategy [8-10], improving the image quality via deep learning [11]. The premise of these studies is that the imaging resolution is global uniform to achieve global high imaging quality. In practice, there is no need to obtain global high-resolution (HR) scenes information. The region of interest (ROI), which is required to be of HR, occupies only a part of the whole field of view (FOV), while the other region of non-interest (NROI) does not need such a HR to provide required information. Thus, achieving HR imaging on ROIs and remaining low-resolution (LR) imaging on NROIs is sufficient to meet practical requirements. Foveated imaging [12], which reduce the data redundancy by sacrificing the resolution of NROIs, is promising to obtain high imaging quality with reduced measurements. The current researches on foveated SPI have proved the feasibility [13-18], but these methods cannot be used directly on under-sampling FSI. Current foveated SPIs mainly realize non-uniform spatial sampling on targets by designing pattern structures with non-uniform resolution. However, non-uniform spatial sampling on the cells (the effective pixels with non-uniform resolution, the number of which is less than the whole pixel with high uniform resolution) cannot obtain the equivalent spatial information, unless the full sampling [13,14] or iterative reconstructing [15-18]. It is not directly applicable in under-sampling Fast Fourier transform (FFT) with information deviation, because the input data of FFT should be tabulated on a uniform grid [19], the non-uniform spatial sampling cannot accurately correspond to the spectrum of cells via FFT at low sampling, especially with strong non-uniformity.

This work reports a uniform-sampling foveated Fourier single-pixel imaging (UFFSI) to achieve under-sampling high-efficiency and high-quality foveated SPI, even in the large-scale scene. First, in order to obtain equivalent spatial information on the non-uniform cells, we add a non-uniform weight distribution (NWD) processing to the foveated pattern structure, being equivalent to achieve uniform sampling. Then, FSI with non-uniform resolution is transformed into FSI with uniform resolution where FFT is used accurately and IFT is used directly to reconstruct the uniform-sampling foveated image with whole cells. In addition, the position selection of ROIs is an important premise for foveated SPI. Fortunately, there are already many mature and efficient object-detection algorithms [20, 21]. It is necessary to combine these methods with the UFFSI for flexible application on various targets. Therefore, the compatibility with object-detection algorithms needs to be considered into the design of foveated pattern structure. Considering the boundary of the ROI and the shape of the target box on current object-detection algorithms, three kinds of foveated pattern structures are designed, namely, "*circular structure*" via log-polar transformation (LPT), "*rectangular structure*" via log-rectilinear transformation (LRT) and "*rotating-rectangle structure*" via log-rectilinear-rotation transformation (LRRT). Note that arbitrary Foveated pattern structures can be flexibly designed according to the needs of targets, as long as each cell and pixel is mapped one by one to ensure the uniform sampling of UFFSI.

## 2. Methods and results

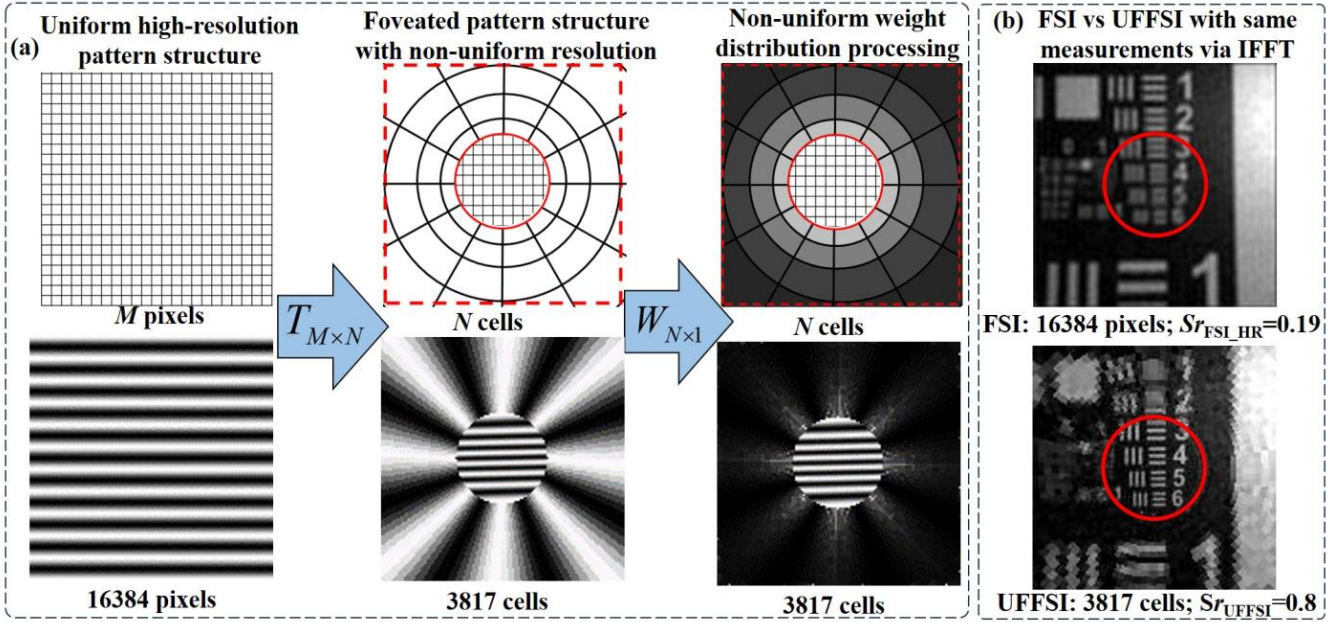

Fig.1. Schematic diagram of UFFSI. (a) Description of foveated Fourier pattern. Taking "*circular structure*" as an example, the foveated pattern structure with total $N$ cells is designed for mapping the uniform HR pattern structure with total $M$ pixels, and then the uniform-sampling foveated Fourier pattern structure is obtained by the NWD processing; where $T_{M \times N}$ is the non-uniform mapping matrix, $W_{N \times 1}$ is the non-uniform weight distribution matrix. (b) Comparison of experimental results between FSI and UFFSI using same under-sampling measurements. Given a foveated Fourier pattern sample with 3817 cells, the imaging quality on the ROI circled in red of UFFSI is much better than the area of uniform HR FSI with 16384 pixels, by reducing the data redundancy of (16384-3817)/16384 = 76.7%. Here, $Sr_{UFFSI}$ is the sampling ratio referring to the total cells of UFFSI, $Sr_{FSI\_HR}$ is the sampling ratio referring to the total pixels of HR FSI.

The schematic diagram of UFFSI is shown in Fig.1. Given a uniform HR pattern structure with total $M$ pixels, according to imaging requirements on ROIs and NROIs, a flexible foveated pattern structure with total $N$ cells is designed via a one-to-one mapping. The foveated pattern structure on the ROI remains such a uniform HR and the area of NROI is designed with non-uniform LR. The mapping relationship is recorded in the non-uniform mapping matrix $T_{M \times N}$, where $M$ is the efficient data number of HR FSI, $N$ is the efficient data number of UFFSI, and $M > N$.

For general FSI with $N$ cells, the mathematic model of Fourier pattern based on FFT is given as [7]:

$$P_{FSI}(n, f_n, \phi) = a + b \cdot \cos(2\pi f_n n + \phi), \tag{1}$$

where $n$ is the $n$th spatial sampling point, $f_n$ is the $n$th sampling frequency, $\phi$ is the initial phase, $a$ is the DC term equal to the average intensity and $b$ is the contrast.

However, the spatial distribution of the $N$ cells on UFFSI is non-uniform. Since the spatial sampling of FFT is required to be uniform [19], Eq. (1) is not directly suitable for generating foveated Fourier pattern with non-uniform spatial sampling, otherwise the corresponding spectral coefficients will not be accurately obtained, thus reducing the imaging quality at under-sampling or strong non-uniformity.

Therefore, in order to obtain accurate spectral coefficients corresponding to the non-uniform $N$ cells, we propose a NWD processing method to make the foveated Fourier pattern be equivalent to be uniform-sampling. According to the non-uniformity on each cell mapping one or more pixels, the non-uniform weight distribution matrix $W_{N \times 1}$ is calculated as:

$$W(n) = \left(\sum_{m=1}^{M} T(m,n)\right)^{-1}, \tag{2}$$

where $W(n)$ is the NWD factor on the $n$th cell, $T(m,n)$ gives the mapping relationship between the $m$th pixel on uniform HR pattern structure and the $n$th cell on foveated pattern structure. $T(m,n)$ =1or 0, where $T(m,n)$=1 means that the $m$th pixel maps the $n$th cell, and vice versa. Then, the uniform-sampling foveated Fourier pattern for UFFSI with $N$ cells is obtained as:

$$P_{UFFSI}(x, y, f_n, \phi) = T \cdot [W(n) \cdot P_{FSI}(n, f_n, \phi)], \tag{3}$$

where $(x,y)$ represents the 2D HR Cartesian coordinates in the scene for $X \times Y = M$ ($X$ is the total pixels in $x$-direction, $Y$ is the total pixels in $y$-direction), $f_n$ is the $n$th uniform-sampling frequency. Note that, there is only $N$ total sampling points due to UFFSI's property of reducing data redundancy.

By illumination the uniform-sampling foveated Fourier patterns on the whole scene, the collected light intensity back-scatted from the scene can be expressed as:

$$S_\phi(f_n) = \sum_{x=1}^{X} \sum_{y=1}^{Y} P_{UFFSI}(x, y, f_n, \phi) \cdot O(x, y), \tag{4}$$

where $O(x,y)$ is the scene on the 2D HR Cartesian coordinates.

Then, using 4-step phase-shifting approach on the UFFSI, the $n$th complex Fourier coefficient $C(f_n)$ can be obtained as:

$$C(f_n) = (S_0(f_n) - S_\pi(f_n)) + j(S_{\pi/2}(f_n) - S_{3\pi/2}(f_n)), \tag{5}$$

where $j$ denotes the imaginary unit. Also, full-sampling of UFFSI needs $2N$ measurements due to conjugate symmetry.

By directly applying the IFT, the accurate non-uniform spatial information of $N$ cells is obtained, and then the image of UFFSI with non-uniform resolution is reconstructed via the non-uniform mapping from $N$ cells to $M$ pixels:

$$O_{UFFSI}(x, y) = T \cdot F^{-1}\{C(f_n)\}, \tag{6}$$

where $F^{-1}$ denotes the IFT.

Using a general experimental setup of FSI (detailed in S1 of Supplement 1), a comparison example between the uniform HR FSI with 128×128 (16384) pixels and the UFFSI with 3817 cells is shown in Fig.1(b). At a low sampling rate of 0.19, the line pairs and numbers are indistinguishable via the HR FSI. But, with the same measurements, both the lines pairs and numbers are clearly distinguishable via the UFFSI with the reduced data redundancy of 76.7%. This shows that the proposed UFFSI can achieve better imaging quality with few measurements than the FSI with uniform-resolution.

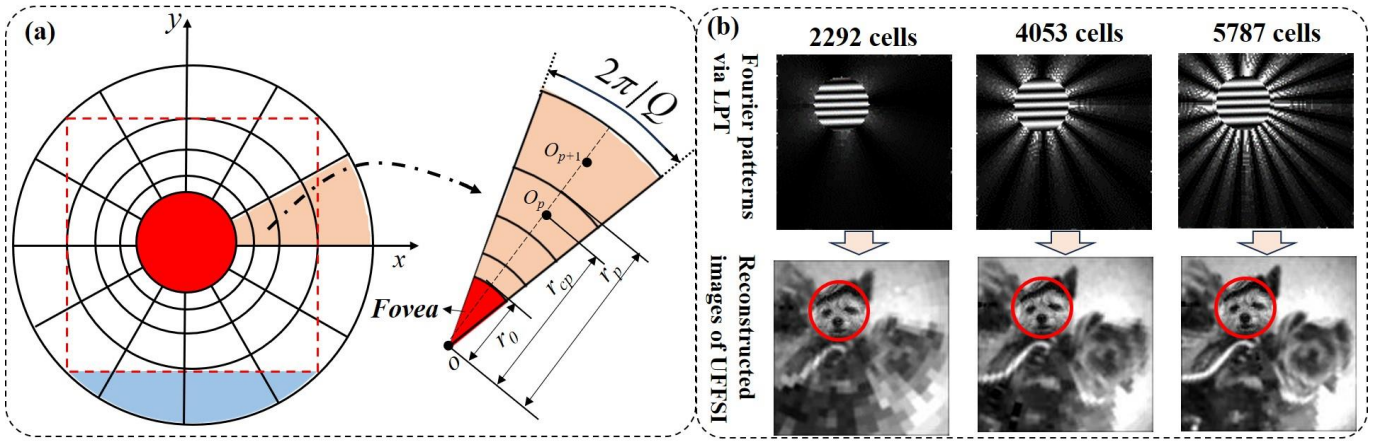

Fig.2. UFFSI with foveated Fourier patterns designing as a "*circular structure*" via LPT. (a) Diagram of "*circular structure*" via LPT. Here, the area framed by the red dotted line is the whole imaging FOV. In order to cover the whole FOV, the maximum radius on the periphery should be the value of the farthest distance from the center point of fovea to the four boundary vertices. Then, the total number of cells on the "*circular structure*" $N \leq (Nc + P \times Q)$ for deleting some cells not in the whole FOV, like the bule area outside of the red dotted line. (b) Foveated Fourier pattern samples via LPT used for UFFSI and the corresponding reconstructed images with $Sr_{UFFSI}$=1. In addition, in this letter, in order to adapt to the human visual system, a simple Gaussian filter is performed on the NROIs of represented images to smooth the nonuniformity of the resolution.

Further, in order to reduce data redundancy to the greatest extent, the boundary of the ROI is an important consideration in the design of foveated Fourier pattern structure. And the shape of the target box on current object-detection algorithms is also considered to accurately obtain the location of the target. Therefore, we flexibly design three kinds of foveated pattern structures, but arbitrary structures can be designed to meet imaging requirements in the real application. The first structure "*circular structure*" via LPT is a classical foveated structure, simulating the distribution of photoreceptor cells in the retina of human eye, as shown in Fig.2(a). The fovea filled in red is circular with a radius $r_0$, the resolution is uniform HR with $Nc$ cells. The periphery filled in orange has $P$ rings with the increasing radius exponentially from the center of the fovea to the outside until covering the whole

FOV, and each ring is evenly divided into $Q$ cells. The mathematical model of the periphery on "*circular structure*" via LPT is shown as:

$$\begin{cases} r_p = r_0 \cdot \varepsilon^p; & r_{c1} = r_0 \cdot \dfrac{1+\varepsilon}{2} \\ \omega_q = q \cdot \dfrac{2\pi}{Q} & (q=1,2,3...Q) \\ \xi_p = \log_\varepsilon(r_{cp}) = \log_\varepsilon(r_{c1}) + p - 1 & (p=1,2,3...P), \end{cases} \quad (7)$$

where $r_p$ is the outer radius of the $p$th ring (the circle with $r_P$ should be cover the whole FOV), $rc_p$ is the center radius of the $p$th ring, $\varepsilon$ is the increasing coefficient of radiuses between adjacent rings, $\omega_q$ is the angle of the $q$th cell of each ring, and $\xi_p$ is the value of the center radius of the $p$th ring in log-polar coordinates.

Due to the limitation of hardware devices, most imaging FOVs are rectangular. Moreover, the common shape of ROIs is rectangular, and horizontal boxes are used frequently for marking targets on the object-detection algorithms. Therefore, in order to better meet the structure requirements, the second structure "*rectangular structure*" via LRT is proposed. Unlike the "*circular structure*" changing only along a radial direction, the non-uniformity of "*rectangular structure*" is designed along both $x$-direction and $y$-direction from the center of the fovea to the outside, which is more flexible on the resolution of NROIs. As shown in Fig.3(a), the fovea filled in red is a rectangle with uniform HR, and the periphery filled in orange is designed with an increasing radius exponentially in the horizontal and vertical directions independently. Given the uniform HR structure with $X \times Y$ pixels and the central point of the ROI $(x_c, y_c)$, in the $x(u)$-$y(v)$ coordinate system of one-to-one mapping, the radius of the ROI in the $x$ direction is $m0$ and that in the $y$ direction is $n0$; the increasing coefficient of radiuses on the NROI in the $x$ direction is $\alpha_1$ and that in the $y$ direction is $\alpha_2$. Then, the mathematical model of "*rectangular structure*" via LRT can be expressed as:

$$\begin{cases} R_1^u = \begin{cases} u, & \text{when } 1 \leq u \leq m0, \\ m0 \cdot \alpha_1^{u-m0}, & \text{when } m0 \leq u \leq (U-1)/2, \end{cases} \\ R_2^v = \begin{cases} v, & \text{when } 1 \leq v \leq n0, \\ n0 \cdot \alpha_2^{v-n0}, & \text{when } n0 \leq v \leq (V-1)/2, \end{cases} \\ x(u) = \begin{cases} \max(1, \text{round}(x_c - R_1^u)), & \text{when } 1 \leq u \leq (U-1)/2, \\ x_c, & \text{when } u = (U+1)/2, \\ \min(\text{round}(x_c + R_1^u), X), & \text{when } (U+1)/2 < u \leq U, \end{cases} \\ y(v) = \begin{cases} \max(1, \text{round}(y_c - R_2^v)), & \text{when } 1 \leq v \leq (V-1)/2, \\ y_c, & \text{when } v = (V+1)/2, \\ \min(\text{round}(y_c + R_2^v), Y), & \text{when } (V+1)/2 < v \leq V, \end{cases} \\ u \in 2N+1, v \in 2N+1 \end{cases} \quad (8)$$

where $R_1^u$ is the radius of the $u$th layer extending from the central point of ROI in the $x$ direction and $R_2^v$ is the radius of the $v$th layer in the $y$ direction. $U$ is the total cells in $x$-direction, $V$ is the total cells in $y$-direction.

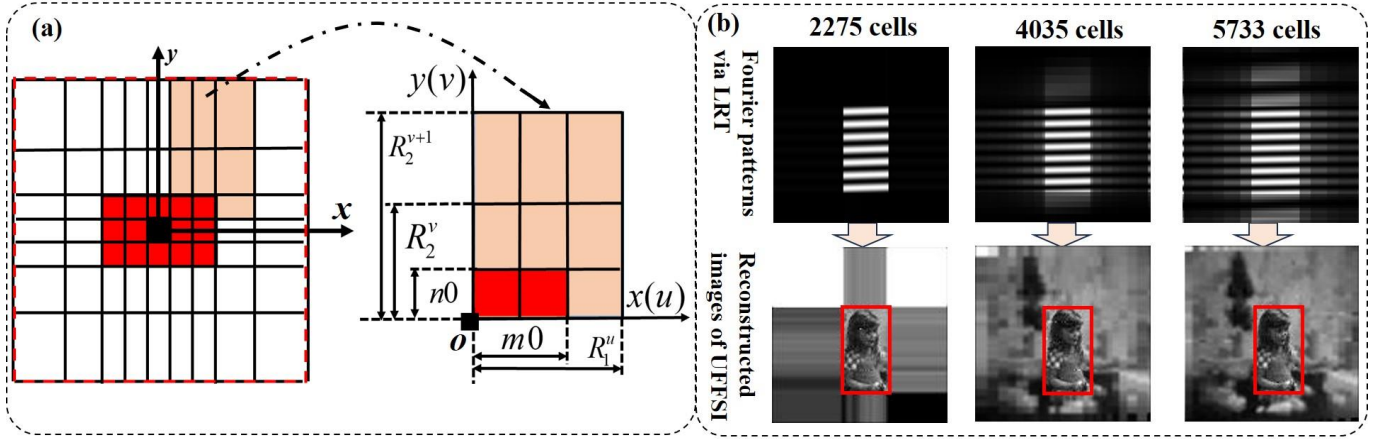

Fig.3. UFFSI with foveated Fourier patterns designing as a "*rectangular structure*" via LRT. (a) Diagram of "*rectangular structure*" via LRT. Here, the total number of cells on the "*rectangular structure*" $N = U \times V$. (b) Foveated Fourier pattern samples via LRT used for UFFSI and the corresponding reconstructed images with $Sr_{UFFSI}=1$.

Further, in some cases, due to the imaging tilt or arbitrary attitude of targets, horizontal boxes in the object-detection algorithm cannot accurately locate targets, similarly, the horizontal rectangle cannot obtain the accurate ROI via the "*rectangle structure*". Thus, inspired by rotating boxes used in object-detection algorithms, the third structure "*rotating-rectangle structure*" via LRRT is proposed based on the "*rectangle structure*" via LRT. Given the rotating angle $\theta$ between the center line of the ROI near $x$-positive direction and $x$-positive direction, the rotated $x'$-$y'$ coordinate system is obtained by using the coordinate rotation transformation via Eq. (9), where the ROI is transformed into a regular rectangle. Note that the coordinate system is rotating counterclockwise around the center point of ROI ($x_c$, $y_c$) and $\theta \in [0,\pi/2)$. Then, in the rotated $x'$-$y'$ coordinate system, with the same principle as the "*rectangle structure*" via Eq. (8), the "*rotating-rectangle structure*" is obtained, as shown in Fig.4(a).

$$\begin{bmatrix} \cos\theta & -\sin\theta \\ \sin\theta & \cos\theta \end{bmatrix} \cdot \begin{bmatrix} x - x_c \\ y - y_c \end{bmatrix} = \begin{bmatrix} x' - x_c \\ y' - y_c \end{bmatrix}. \tag{9}$$

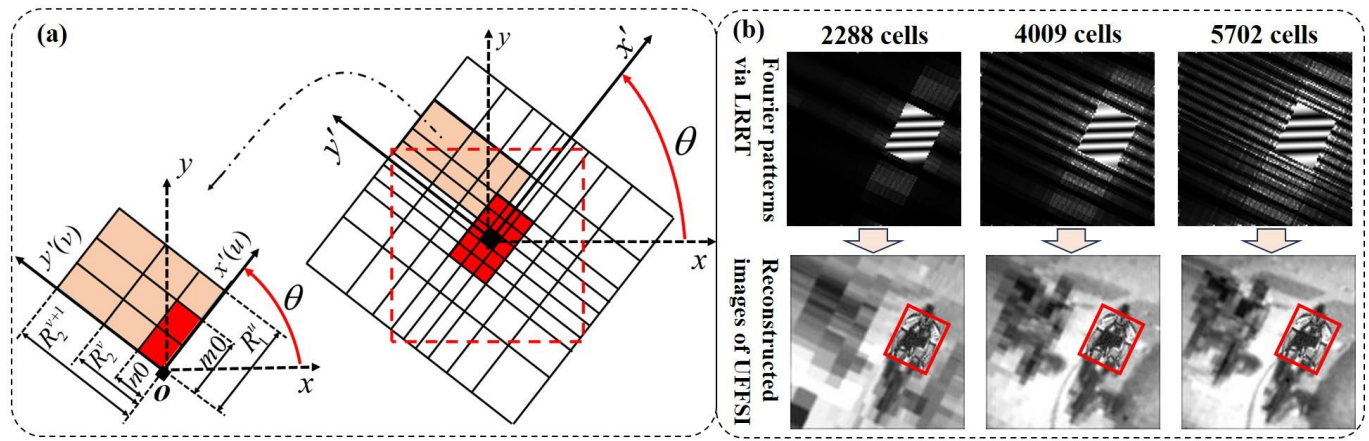

Fig.4. UFFSI with foveated Fourier patterns designing as a "*rotating-rectangle structure*" via LRRT. (a) Diagram of "*rotating-rectangle structure*" via LRRT. Here, in the rotated $x'$-$y'$ coordinate system with a rotating angle $\theta$, the whole designed area should be larger than the original area to cover the whole FOV. Then, the total number of cells on the "*rotating-rectangle structure*" $N \leq U \times V$ for deleting some cells not in the whole FOV, with the same reason as the "*circular structure*". (b) Foveated Fourier pattern samples via LRRT used for UFFSI and the corresponding reconstructed images with $Sr_{UFFSI}=1$.

All the structures are designed to accurately locate ROIs to avoid additional data redundancy, and further meet imaging requirements for ROIs and NROIs with fewer measurements. Therefore, according to different imaging requirements for ROIs and NROIs, by selecting appropriate foveated pattern structure and designing corresponding parameters, flexible foveated Fourier patterns are generated for completing UFFSI.

Combining the principle and experimental verification, we summarize the three features of the proposed UFFSI, namely, uniform sampling, effective sampling, and flexible fovea. The first feature of UFFSI, uniform sampling, is reflected in that the uniform spatial-information sampling of $N$ cells with non-uniform spatial distribution. Given a fixed ROI, the spatial information collected by $N_c$ cells corresponding to the ROI is similar, and finally the imaging effect in the ROI is similar, as shown in Fig.2-Fig.4(b) (detail in S2 of Supplement 1). At this case, the size of $N$ depends on the reduction degree of data redundancy on the NROI. If the imaging on NROI is required to be clear, $N$ is large; if there is little imaging requirement on NROI, the size of $N$ should be reduced as much as possible to reduce data redundancy, so as to obtain high-quality imaging of ROI with fewer sampling measurements. Therefore, from this feature, parameters of foveated pattern structure can be determined according to information requirements of NROIs.

The second feature, effective sampling, is reflected in that under the same sampling measurements, the imaging quality on ROI of UFFSI is better than that of FSI with same uniform data number and that of FSI with the same uniform HR (detail in S3 of Supplement 1). More importantly, the proposed UFFSI is also effective for large-scale imaging. To be our best knowledge, foveated imaging has not yet been used to achieve high-quality SPI on large-scale scene, due to the little high time cost caused by full sampling or iterative reconstructing of current Foveated SPIs. However, using UFFSI, the large-scale scene with 1024×768 pixels (maximum resolution that can be achieved with the used DMD) is experimentally reduced into 255×341 cells with the reduced data redundancy of 89%, and even more according to imaging requirements. As shown in Fig.5, with similar measurements at an ultralow sampling rate of 0.0084 referring to the uniform HR FSI, the ROI boxed in red on UFFSI is clear with discernible line pairs and numbers, while the ROI on FSI with uniform HR or similar data number is blurry with a bad imaging quality. This proves that the UFFSI is an effective way to achieve high imaging quality and high imaging efficiency on SPI.

The third feature, flexible fovea, is reflected in the number and position of ROIs. Due to the arbitrariness of ROIs in the real application, three flexible foveated pattern structures are proposed. Experimental results of multi-fovea UFFSI in S4 of Supplement 1 prove the flexibility of the proposed foveated pattern structures on various ROIs of different nature scenes. Further, foveated pattern structures, including but not limited to the proposed three structures are designed flexibly for UFFSI to meet the imaging requirements, as long as non-uniform cells and uniform HR pixels are mapped one by one.

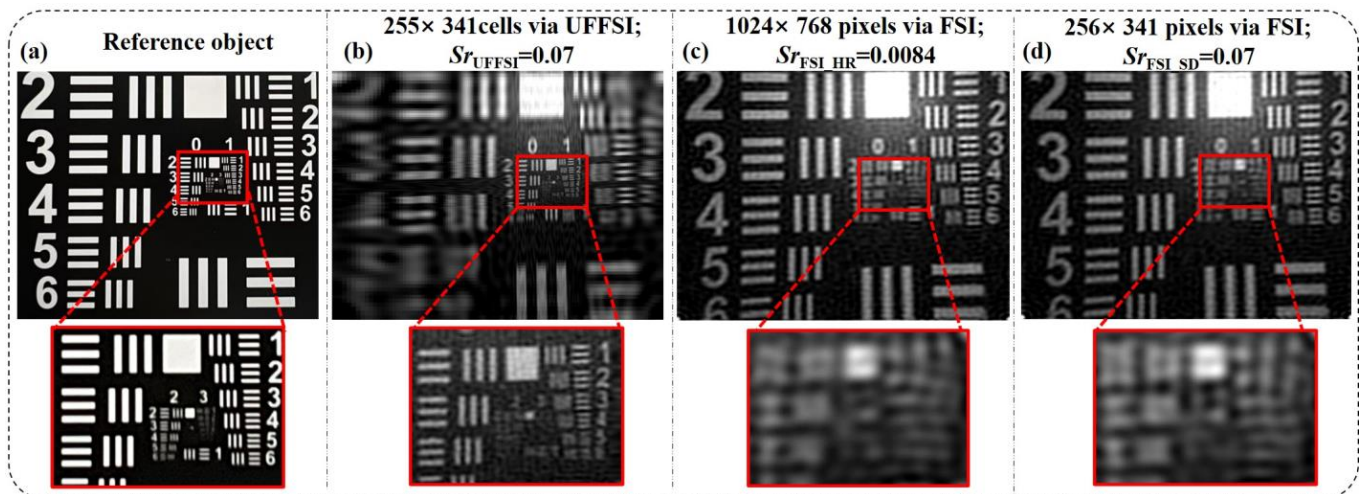

Fig.5. Large-scale UFFSI. (a) Reference object. (b) UFFSI with 255×341 cells at the sampling ratio of 0.07 referring to 255×341=86955 cells with 12172 measurements. (c) FSI with uniform 1024×768 pixels at the sampling ratio of 0.0084 referring to 1024×768=786432 cells with 13212 measurements. (d) FSI with 256×341 pixels at the sampling ratio of 0.07 referring to 256×341=87296 cells with 12220 measurements.

## 3. Conclusion

In summary, the proposed UFFSI can obtain high imaging quality with few measurements via three features of uniform sampling, effective sampling and flexible fovea, even in large-scale scene. First, by flexibly using the three foveated pattern structures, data redundancy is reduced significantly to only require HR on accurate ROIs and remain LR on NROI, which is an effective way radically reducing the need of total data number. Next, by the NWD processing, non-uniform spatial sampling is transformed into uniform sampling, and FFT is used accurately and directly to obtain under-sampling high imaging quality with further reduced measurements. We hope this work can provide a breakthrough for real-time SPI.